\documentclass[Physsubmission, Phys]{SciPost}
\hypersetup{
    colorlinks,
    linkcolor={red!50!black},
    citecolor={blue!50!black},
    urlcolor={blue!80!black}
}

\usepackage[bitstream-charter]{mathdesign}
\usepackage[T1]{fontenc}
\usepackage{bm}
\usepackage{graphicx}
\usepackage{lscape}
\usepackage{subfig}

\DeclareMathOperator{\dlog}{\mathit{d}log}

\usepackage{mathtools}

\DeclareSymbolFont{usualmathcal}{OMS}{cmsy}{m}{n}
\DeclareSymbolFontAlphabet{\mathcal}{usualmathcal}

\begin{document}

\begin{center}{\Large \textbf{
Towards NNLO corrections
to muon-electron scattering\\
}}\end{center}

% TODO: write the author list here. Use initials + surname format.
% Separate subsequent authors by a comma, omit comma at the end of the list.
% Mark the corresponding author with a superscript *.
\begin{center}
Syed Mehedi Hasan\textsuperscript{1$\star$}
\end{center}

\begin{center}
{\bf 1} INFN, Sezione di Pavia, Via A. Bassi 6, 27100 Pavia, Italy
\\
* syedmehe@pv.infn.it
\end{center}

\begin{center}
\today
\end{center}

% For convenience during refereeing (optional),
% you can turn on line numbers by uncommenting the next line:
%\linenumbers

\definecolor{palegray}{gray}{0.95}
\begin{center}
\colorbox{palegray}{
  \begin{tabular}{rr}
  \begin{minipage}{0.1\textwidth}
    \includegraphics[width=35mm]{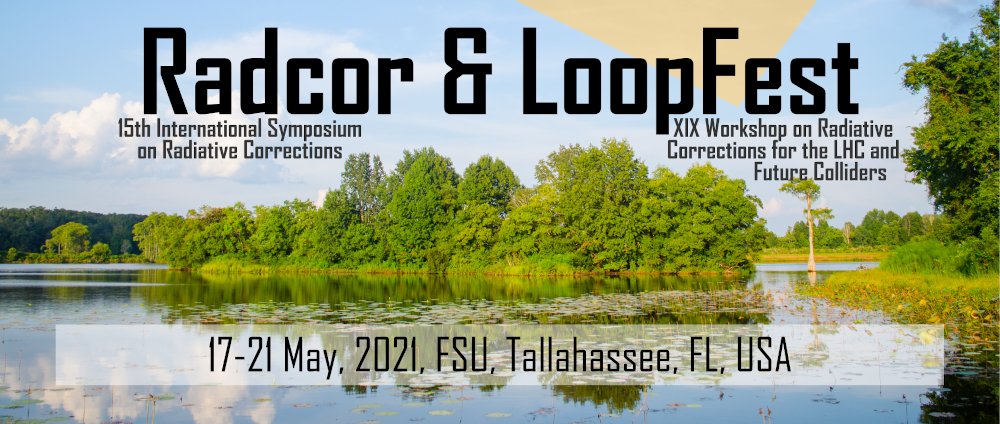}
  \end{minipage}
  &
  \begin{minipage}{0.85\textwidth}
    \begin{center}
    {\it 15th International Symposium on Radiative Corrections: \\Applications of Quantum Field Theory to Phenomenology,}\\
    {\it FSU, Tallahasse, FL, USA, 17-21 May 2021} \\
    \doi{10.21468/SciPostPhysProc.?}\\
    \end{center}
  \end{minipage}
\end{tabular}
}
\end{center}

\section*{Abstract}
{\bf
% TODO: write your abstract here.
The recently proposed MUonE experiment at CERN aims at providing a novel determination of the leading order hadronic contribution to the muon anomalous magnetic moment through the study of elastic muon-electron scattering at relatively small momentum transfer. The anticipated accuracy of the order of 10ppm demands for high-precision predictions, including all the relevant radiative corrections. To aid the effort, the theoretical formulation for the fixed order NNLO QED corrections are described and the virtual NNLO leptonic corrections are calculated with complete mass effects.~\footnote{* speaker}
}

\vspace{10pt}
\noindent\rule{\textwidth}{1pt}
\tableofcontents\thispagestyle{fancy}
\noindent\rule{\textwidth}{1pt}
\vspace{10pt}

\section{Introduction}
\label{sec:intro}
% TODO: write your article here.
The anomalous magnetic moment of the muon, $a_\mu =
(g-2)_\mu / 2$, is a fundamental quantity in particle physics. This quantity has been recently measured by the Fermilab Muon $g-2$ Experiment
(E989) with relative precision of
$0.46$ppm~\cite{Abi:2021gix} for positive muons, which is an improvement with respect to the
0.54ppm precision obtained by the E821 experiment at the Brookhaven
National Laboratory~\cite{Bennett:2006fi}.

The muon anomaly is due to loop corrections
coming from the QED, weak and strong sector of the
Standard Model (SM)~\cite{Jegerlehner:2009ry,Jegerlehner:2017gek}. There is a long-standing puzzle that the discrepancy between the measured value and theoretical value of $g-2$ currently
exceeds $3\sigma$ level. Morover it
grows up to $4.2\sigma$ when combined with previous measurements using both $\mu^+$ and $\mu^-$
beams~\cite{Abi:2021gix}. The present status of the SM theoretical
prediction has been recently reviewed in
ref.~\cite{Aoyama:2020ynm}.

The E34
experiment, which is under development at
J-PARC~\cite{Iinuma:2011zz,Mibe:2011zz} and future runs of the E989 experiment at
Fermilab~\cite{Venanzoni:2014ixa,Grange:2015fou}, are expected to bring the
relative precision below
the level of $0.2$ppm.  Naturally a major effort from the theory
side is called for to reduce the uncertainty in the SM prediction, most of which
comes from the non-perturbative strong interaction effects as given by
the leading order hadronic correction, $a_\mu^\text{HLO}$, and the
hadronic light-by-light contribution. There is at present
general consensus that the latter has a small impact on the
theoretical uncertainty on $a_\mu$~\cite{Aoyama:2020ynm}.

The traditional computation of $a_\mu^\text{HLO}$ is performed via dispersion
integral of the hadron production cross section in electron-positron
annihilation at low
energies~\cite{Jegerlehner:2018zrj,Keshavarzi:2018mgv,Davier:2019can,Aoyama:2020ynm}. An alternative evaluation of the $a_\mu^\text{HLO}$ can be carried out using Lattice QCD calculation~\cite{DellaMorte:2017dyu,Chakraborty:2017tqp,Borsanyi:2017zdw,Meyer:2018til,Blum:2018mom,Giusti:2018mdh,Giusti:2019xct,Giusti:2019hkz,Shintani:2019wai,Davies:2019efs,Gerardin:2019rua,Aubin:2019usy,Giusti:2020efo,Lehner:2020crt}. Recently the calculation carried out by the BMW collaboration shows a more precise determination of $a_\mu^\text{HLO}$ ~\cite{Borsanyi:2020mff} with a central value larger than the dispersive approach and in closer agreement with BNL and FNAL ~\cite{Bennett:2006fi,Abi:2021gix} experiment outcome. To
resolve the tension, alternative approaches for the evaluation of $a_\mu^\text{HLO}$ are more than welcome.

A novel approach to derive $a_\mu^\text{HLO}$ from a measurement of the effective
electromagnetic coupling constant in the space-like region via
scattering data, has been proposed in ref.~\cite{Calame:2015fva}.
The elastic scattering of high-energy muons on atomic electrons has been
identified as an ideal process ~\cite{Abbiendi:2016xup}
and as a result a new experiment, MUonE, has been proposed at CERN~\cite{MUonE:LoI,Abbiendi:2019qtw,Ballerini:2019zkk,Abbiendi:2021xsh}.
This new determination of $a_\mu^\text{HLO}$ can be
competitive with the dispersive approach, if the uncertainty
in the measurement of the $\mu e$ differential cross section is
of the order of 10ppm. This challenge of MUonE requires very high-precision
predictions for $\mu e$ scattering, which includes all the relevant
radiative corrections, implemented in fully-fledged Monte Carlo (MC)
tools. The necessary perturbative accuracy is at next-to-next-to
leading order in QED (NNLO QED), combined with the leading logarithmic contributions due to multiple
photon radiation. A comprehensive review of the recent developments along this line, as of the beginning of 2020, can be found in
ref.~\cite{Banerjee:2020tdt}. The complete calculation at NLO accuracy and the
        evaluation of its phenomenological impact on MUonE observables
        can be found in
ref.~\cite{Alacevich:2018vez}. At NNLO accuracy there has been significant progress recently~\cite{Mastrolia:2017pfy,DiVita:2018nnh,Engel:2018fsb,Engel:2019nfw,CarloniCalame:2020yoz,Banerjee:2020rww,Bonciani:2021okt,Fael:2018dmz,Fael:2019nsf,Heller:2021gun,Budassi:2021twh}.~\footnote{It is worth reminding the studies of refs.~\cite{Dev:2020drf,Masiero:2020vxk}
        which investigate the robustness of the MUonE measurements
        to possible New Physics effects}

In the following we summarize our effort along NNLO photonic corrections and NNLO leptonic corrections to muon-electron scattering, which are the subject matter of ref.~\cite{CarloniCalame:2020yoz} and ref.~\cite{Budassi:2021twh} respectively.

\section{NNLO photonic corrections to muon-electron scattering}

\begin{figure}[hbtp]
\begin{center}
\includegraphics[width=0.95\textwidth]{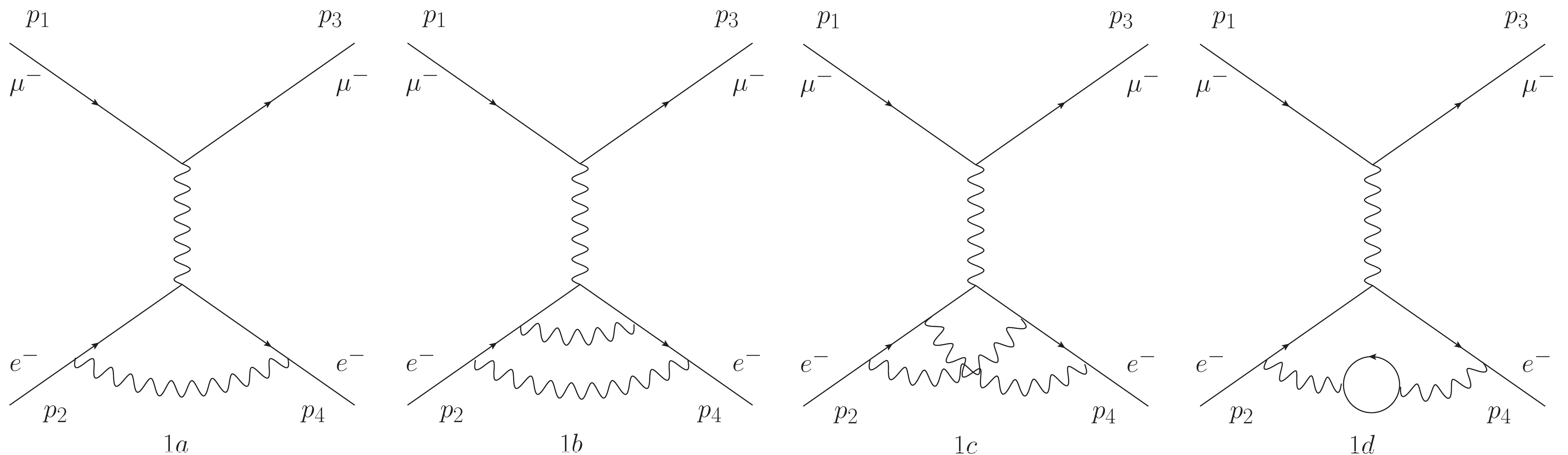}
\end{center}
\caption{Virtual QED corrections to the electron line 
in $\mu e \to \mu e$ scattering. One-loop correction (diagram 1a); 
sample topologies for the two-loop corrections (diagrams 1b-1d). 
The blob in diagram 1d denotes an electron loop insertion. On-shell 
scheme counterterms are understood.}
\label{fig:fig1}
\end{figure}

We define NNLO photonic corrections by removing the contribution
due to the fermionic loop (for example, in Figure~\ref{fig:fig1}, diagram 1d is removed), which forms a gauge invariant subset. Further justification is shown in favour of a stand-alone analysis for this type of contribution in ref.~\cite{CarloniCalame:2020yoz}. All the calculations are exact except the double-box like diagrams, where YFS inspired approximation ~\cite{Yennie:1961ad} is used to capture the correct IR structure. In other words, the
approximation misses only the non-IR remnant of
the two-loop diagrams where at least two photons connect the
electron and muon lines.

\begin{figure}[hbtp]
\begin{center}
\includegraphics[width=0.95\textwidth]{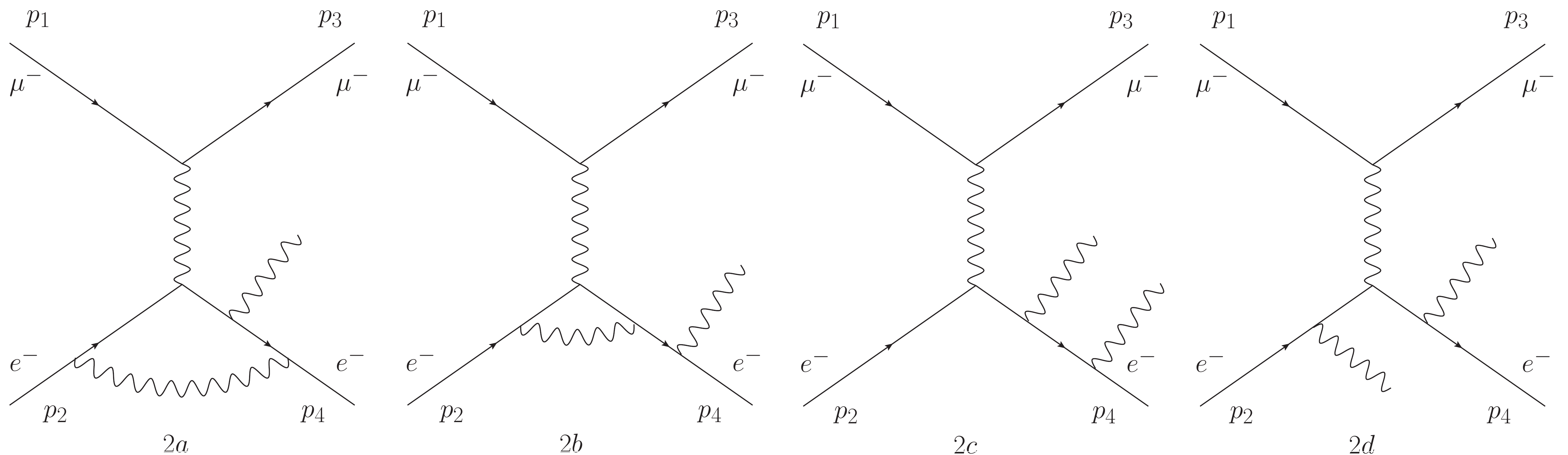}
\end{center}
\caption{Sample diagrams for the one-loop QED corrections to single photon emission (diagrams 2a-2b); 
sample diagrams for the double bremsstrahlung process (diagrams 2c-2d).}
\label{fig:fig2}
\end{figure}

We studied the numerical impact of QED NNLO photonic corrections to observables
assuming the following event selections
(Setup~2 and Setup~4 of Ref.~\cite{Alacevich:2018vez}): 
\begin{enumerate}
\item  $\theta_e,\theta_\mu<100\text{\ mrad}$ and $E_e > 1$~GeV
  {(i.e. $t_{ee}\lesssim -1.02\cdot 10^{-3}\text{ GeV}^2$)}. The angular cuts model the typical acceptance 
conditions of the experiment and the electron energy threshold is imposed to
guarantee the presence of two charged tracks in the detector (Setup~1);
\item the same criteria as above, with the additional acoplanarity cut 
  $|\pi - |\phi_e-\phi_\mu||\le 3.5\text{\ mrad}$. This event selection is considered in order to mimic an experimental
  cut which allows to stay close to the elasticity curve given by the
  tree-level relation between the electron and muon scattering angles
  (Setup~2). 
\end{enumerate}  

The input parameters for the simulations are set to
\begin{equation}
  \alpha = 1/137.03599907430637 \qquad 
  m_e = 0.510998928\text{ MeV} \qquad m_\mu = 105.6583715\text{ MeV}\nonumber
  \label{eq:inputqedparams}
\end{equation}
and the considered differential observables are the following ones:
\begin{equation}
  \frac{d\sigma}{d t_{ee}}\,, \, \, \, \, \, \, 
  \frac{d\sigma}{d t_{\mu \mu}}\,, \, \, \, \, \, \,
  \frac{d\sigma}{d\theta_e}\,, \, \, \, \, \, \,
  \frac{d\sigma}{d\theta_\mu}\,, \, \, \, \, \, \,
  \frac{d\sigma}{d E_e}\,, \, \, \, \, \, \,
  \frac{d\sigma}{d E_\mu} \,,
  \label{eq:observables}
  \end{equation}
where $t_{ee} = (p_2 - p_4)^2$, $t_{\mu \mu} = (p_1 - p_3)^2$,
$(\theta_e, E_e)$ and $(\theta_\mu, E_\mu)$ are the scattering angle 
and the energy, in the laboratory frame, of the outgoing electron and muon,
respectively.

At this point it is useful to define two quantities as follows:
\begin{equation}
  \Delta_\text{NNLO}^i \, = \,
  100 \times \frac{d\sigma_\text{NNLO}^i - d\sigma_\text{NLO}^i}{d\sigma_\text{LO}}\, , 
  \label{eq:delta-def}
\end{equation}
where $i = e, \mu, f$ implies electron-line only, muon-line only
and the full set of NNLO corrections.
Similarly we define:
\begin{equation}
  \Delta_\text{NLO}^i \, = \,
  100 \times \frac{d\sigma_\text{NLO}^i - d\sigma_\text{LO}}{d\sigma_\text{LO}}\, .
  \label{eq:delta_NLO-def}
\end{equation} 

\begin{figure}[t]
\begin{center}
\includegraphics[width=0.95\textwidth]{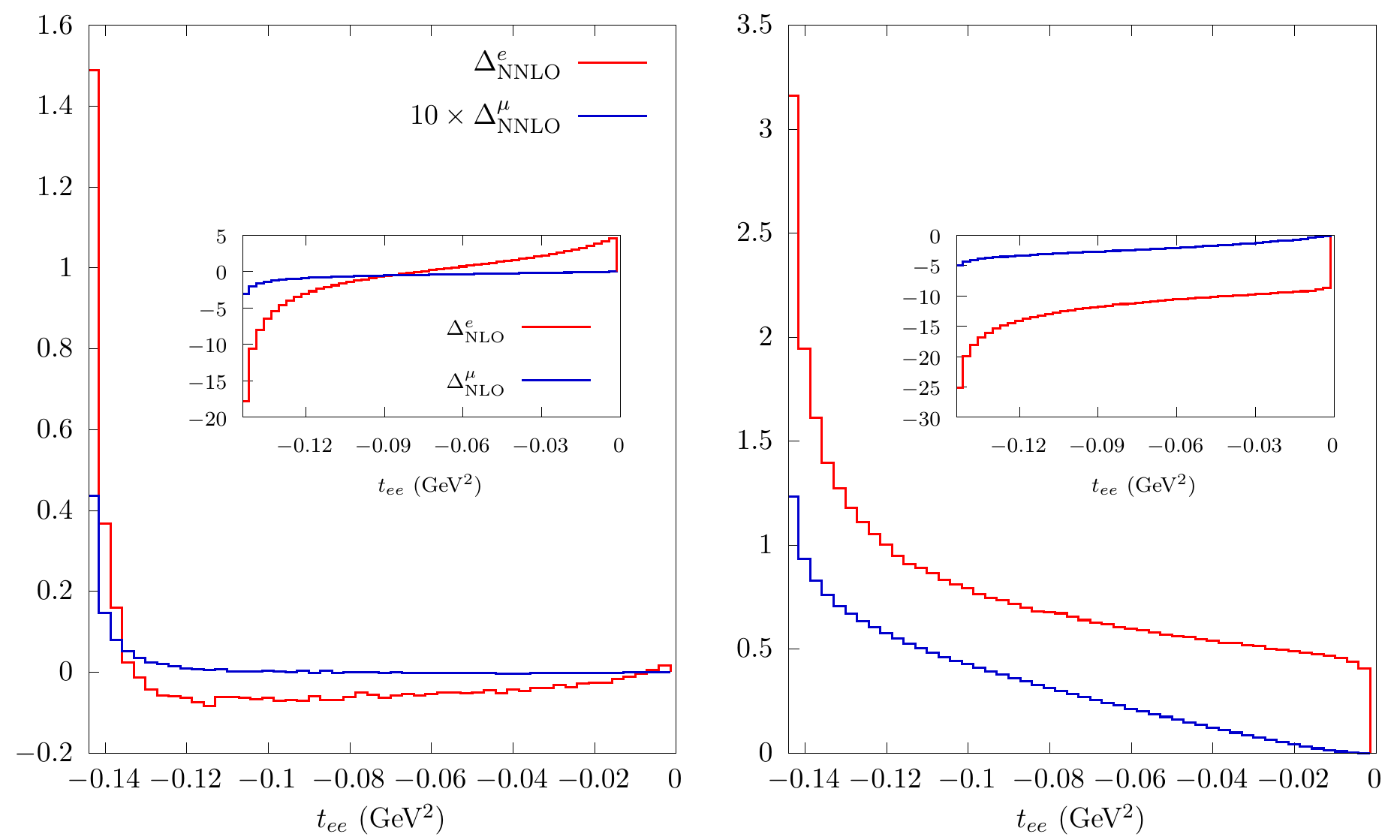}
\end{center}
\caption{\label{Fig:t24-0-1} Red histograms: NNLO corrections, 
  according to Eq.~(\ref{eq:delta-def}), along the electron line,
  for the $\mu^\pm e^- \to \mu^\pm e^- $ process, as a function of
  the squared momentum transfer measured along the electron line, $t_{ee}$;
  blue histogram: $10\times$NNLO corrections along the muon line. 
  Left panel: differential distribution in the presence of acceptance cuts
  only (Setup~1); right panel: the same as in the left panel with
  the additional acoplanarity cut of Setup~2.
  The insets report the NLO corrections,
  according to Eq.~(\ref{eq:delta_NLO-def}); the blue line represents
  directly Eq.~(\ref{eq:delta_NLO-def}) without any multiplicative factor.}
\end{figure}

\begin{figure}[t]
\begin{center}
\includegraphics[width=0.95\textwidth]{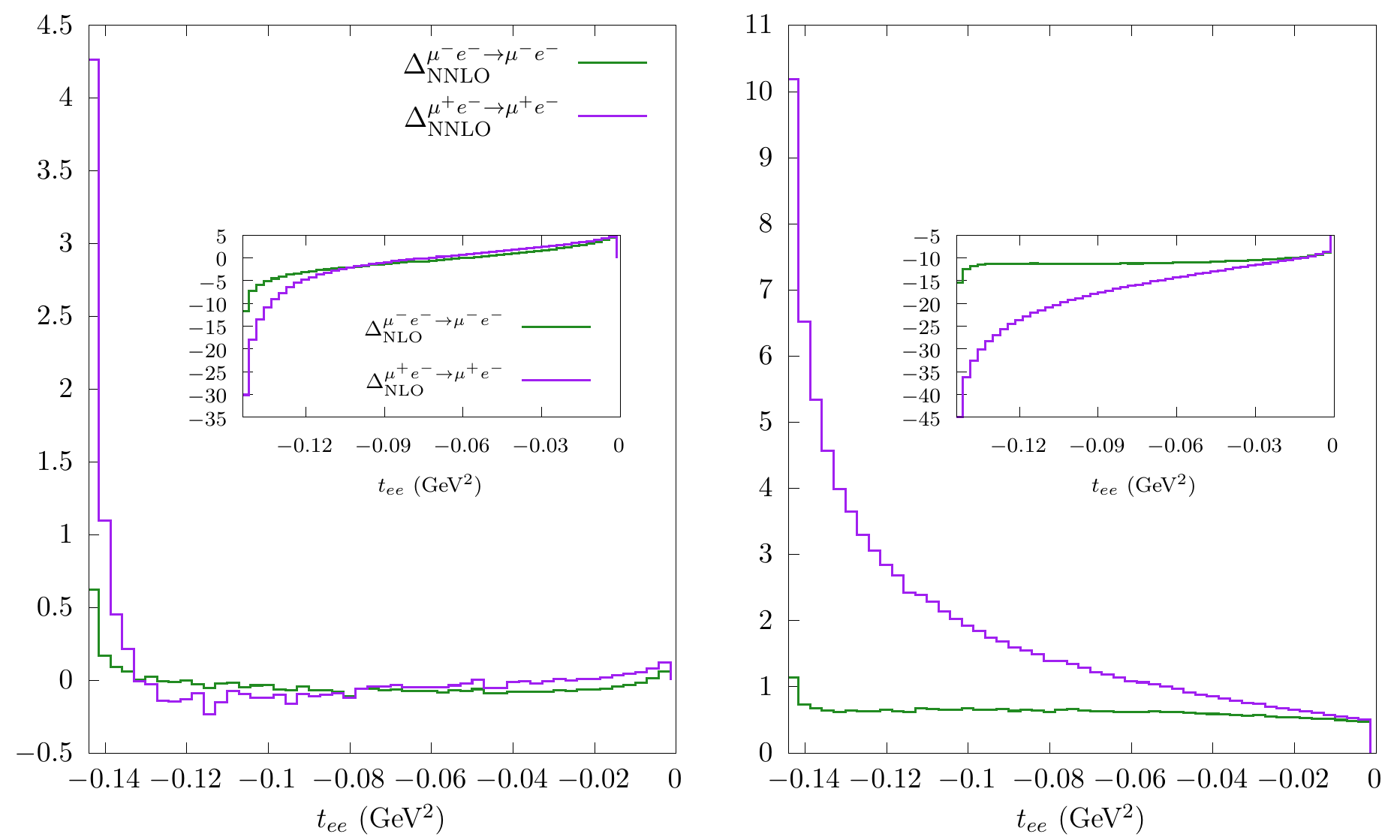}
\end{center}
\caption{\label{Fig:t24-int} Approximation (based on YFS) to the full NNLO corrections 
  as a function of the squared momentum transfer measured along the electron line, $t_{ee}$.
  Green histogram refers to the process $\mu^- e^- \to \mu^- e^- $ while 
  the violet histogram refers to $\mu^+ e^- \to \mu^+ e^- $. 
  Left panel: differential distribution in the presence of acceptance cuts
  only (Setup~1); right panel: the same as in the left panel
  with the additional acoplanarity cut of Setup~2.
  The insets report the NLO corrections, according to
  Eq.~(\ref{eq:delta_NLO-def}).}
\end{figure}

Fig.~\ref{Fig:t24-0-1} shows relative contribution, differential in the squared momentum
         transfer measured along the electron line, of the exact NNLO corrections from one leptonic line. Whereas Fig.~\ref{Fig:t24-int} shows the relative contribution of the full(approximate) NNLO corrections. Similar analysis are performed with respect to the other listed observables ~\cite{CarloniCalame:2020yoz}.  

\section{Vertex and box-like irreducible double virtual NNLO leptonic corrections to muon-electron scattering}
\begin{figure}[hbtp]
        \begin{center}
          \includegraphics[width=0.23\textwidth]{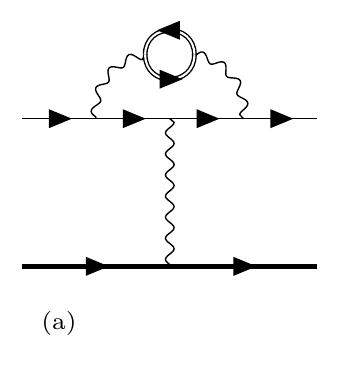}\hspace{0.5cm}
          \includegraphics[width=0.23\textwidth]{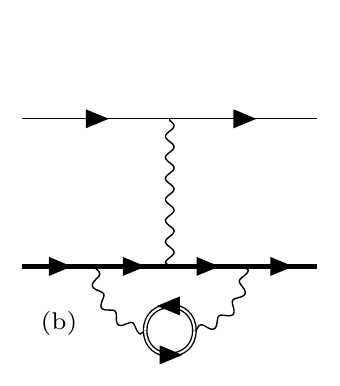}
        \caption{
          NNLO irreducible vertex diagrams. Vertex correction
          along the electron line (left) and along the muon line (right).}
\label{Fig:irred-vert}
\end{center}
\end{figure}

\begin{figure}[hbtp]
        \begin{center}
          \includegraphics[width=0.23\textwidth]{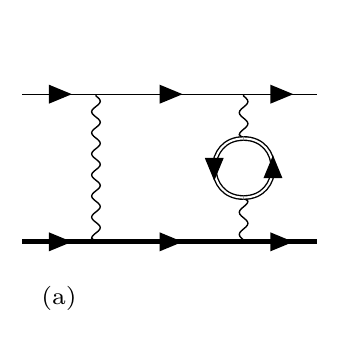}\hspace{0.1cm}
          \includegraphics[width=0.23\textwidth]{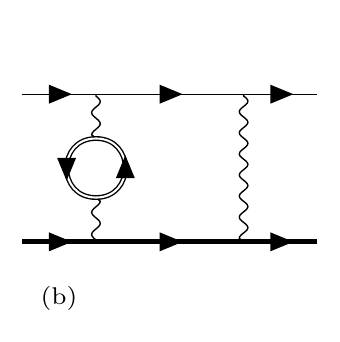}\hspace{0.1cm}
          \includegraphics[width=0.23\textwidth]{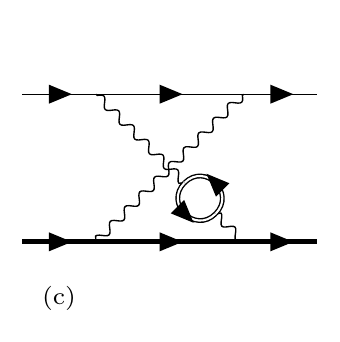}\hspace{0.1cm}
          \includegraphics[width=0.23\textwidth]{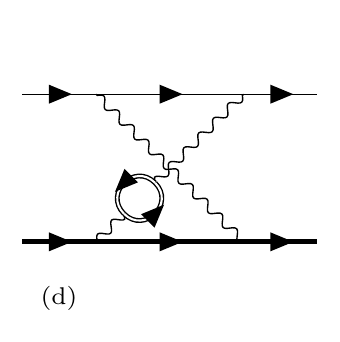}
        \caption{
          NNLO Box diagrams. Direct diagrams (a)-(b) and 
          crossed diagrams (c)-(d).}
\label{Fig:irred-box}
\end{center}
\end{figure}

The calculation of the complete fixed-order
NNLO QED corrections which include at least one leptonic pair, with
all the relevant virtual and real contributions, summing over all
contributing lepton flavours, is carried out in ref.~\cite{Budassi:2021twh}. We name this class of contribution as the NNLO leptonic corrections. Here we only report the vertex and box-like irreducible double virtual case, represented by Fig.~\ref{Fig:irred-vert} and \ref{Fig:irred-box} respectively. 

\begin{figure}[hbtp]
\begin{center}
\includegraphics[width=0.8\textwidth]{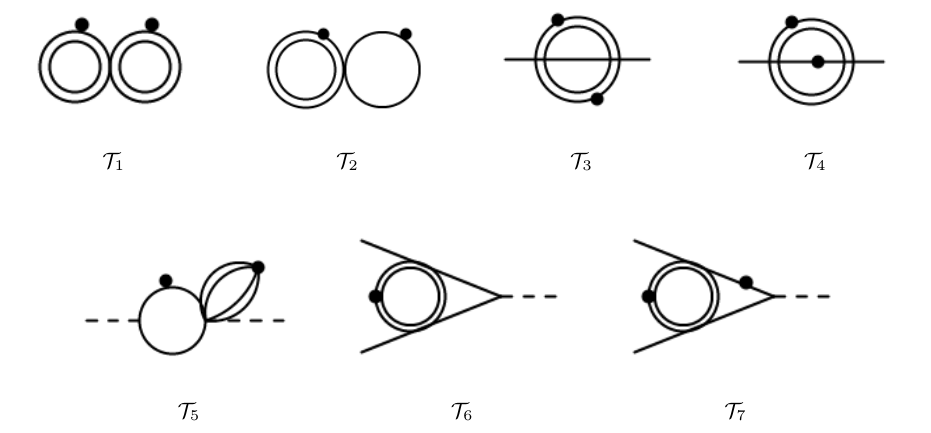}
\end{center}
\caption{The master integral topologies $\mathcal{T}_{1\dots7}$ for
   fermionic loop induced NNLO vertex correction.  The solid single
   line represents a propagator with mass $m_1$ and the double line
   represents a propagator with mass $m_2$, while the dashed line
   represents a massless propagator with momentum squared equal to
   $t$. The dots represent additional powers of the propagator. }
\label{fig:MIs}
\end{figure}

We use dispersion relation (DR) technique to calculate the above mentioned contributions. An
amplitude involving a photon in a loop with a vaccuum polarisation (VP) insertion, due to
lepton $\ell$, can be extracted by replacing the photon propagator~\cite{Hoang:1995ex,Actis:2007fs,Actis:2008sk,Kuhn:2008zs,Barbieri:1972as,Barbieri:1972hn} as
follows
\begin{equation}
  \frac{-i g_{\mu \nu}}{q^2 + i \epsilon} \to
  \frac{-i g_{\mu \delta}}{q^2 + i \epsilon}
  i\left(q^2 g^{\delta \lambda} - q^\delta q^\lambda \right)
  \Pi_\ell(q^2)\frac{-i g_{\lambda \nu}}{q^2 + i \epsilon}\, ,
  \label{Eq:photon-prop-subst}
\end{equation}
where the renormalised $\Pi(q^2)$ is obtained
from its imaginary part using the subtracted DR
\begin{equation}
  \Pi_\ell(q^2)= -\frac{q^2}{\pi}\int_{4 m_\ell^2}^{\infty} \frac{dz}{z}
  \frac{\operatorname{Im}\;\Pi_\ell(z)}{q^2 - z + i \epsilon}\,
  \label{Eq:subtrDR}
\end{equation}
where
\begin{eqnarray}
  \text{Im}\Pi_\ell(z) &=& - \frac{\alpha}{3} R_\ell(z) \, , \\
  R_\ell(z) &=& \left( 1 + \frac{4 m_\ell^2}{2 z}\right)
  \sqrt{1 - \frac{4 m_\ell^2}{z}} 
  \label{{Eq:leptonic-Pi}}\, ,
\end{eqnarray}
$m_\ell$ representing the mass of the lepton blob.
The $q^\delta q^\lambda$ term in eq.~(\ref{Eq:photon-prop-subst})
does not contribute because of gauge invariance. Hence the two-loop
calculation can be performed starting from the one-loop
amplitude with the replacement
\begin{equation}
  \frac{-i g_{\mu \nu}}{q^2 + i \epsilon} \to - i g_{\mu \nu} \left(
  \frac{\alpha}{3 \pi} \right) \int_{4 m_\ell^2}^\infty \frac{dz}{z}
  \frac{1}{q^2 - z + i \epsilon}\left( 1 + \frac{4 m_\ell^2}{2 z}\right)
  \sqrt{1 - \frac{4 m_\ell^2}{z}}\, .
  \label{Eq:leptonic-subtrDR}
\end{equation}

Furthermore, we have analyticaly evaluated the vertex form factors (Fig.~\ref{Fig:irred-vert}) by evaluating master integrals (MI) and found agreement with the DR method.

Seven master integrals, depicted in figure~\ref{fig:MIs}, are evaluated. We used differential equation
methods~\cite{Kotikov:1990kg,Remiddi:1997ny,Gehrmann:1999as} and chose
a suitable basis such that the dimensional
regularisation parameter $\epsilon$ factorises from the kinematics. These
integrals were then encoded in a $\dlog$-form, also called canonical
form~\cite{Henn:2013pwa}, which we obtained using the
Magnus algorithm~\cite{DiVita:2014pza,Argeri:2014qva}. 
Finally the master integrals are written in terms of the generalised
polylogarithms~\cite{Remiddi:1999ew,Gehrmann:2001jv,Vollinga:2004sn}. The explicit results are available in the ancillary files of the \texttt{arXiv}
version of ref.~\cite{Budassi:2021twh}.

\section{Conclusion}
We have analyzed the NNLO photonic 
corrections to $\mu^\pm e^- \to \mu^\pm e^-$ scattering. The phenomenological results are obtained with a related MC code,
named \textsc{Mesmer}. 
The motivation of the study is due to the recent proposal of measuring the effective electromagnetic coupling constant in the space-like region by using this particular process
(MUonE experiment). From the measurement of the 
hadronic contribution to the running of $\alpha_\text{QED}$, the leading hadronic contribution to the muon anomaly 
can be derived as an alternative approach to 
the standard time-like evaluation of $a_\mu^\text{HLO}$.

The NNLO leptonic virtual contribution have been calculated with dispersion
relation techniques including all finite mass effects. For the class
of vertex corrections, the numerical results have been cross-checked
with exact analytical result, valid for arbitrary internal and
external masses, finding excellent agreement.

\section*{Acknowledgements}
We are grateful to the MUonE colleagues for many useful discussions. SMH acknowledges INFN Sezione di Pavia for providing the necessary funding and hospitality to carry out the research.

% TODO:
% Provide your bibliography here. You have two options:

% FIRST OPTION - write your entries here directly, following the example below, including Author(s), Title, Journal Ref. with year in parentheses at the end, followed by the DOI number.
%\begin{thebibliography}{99}
%\bibitem{1931_Bethe_ZP_71} H. A. Bethe, {\it Zur Theorie der Metalle. i. Eigenwerte und Eigenfunktionen der linearen Atomkette}, Zeit. f{\"u}r Phys. {\bf 71}, 205 (1931), \doi{10.1007\%2FBF01341708}.
%\bibitem{arXiv:1108.2700} P. Ginsparg, {\it It was twenty years ago today... }, \url{http://arxiv.org/abs/1108.2700}.
%\end{thebibliography}

% \bibliographystyle{SciPost_bibstyle} % Include this style file here only if you are not using our template
\bibliographystyle{SciPost_bibstyle}
\bibliography{muone_pairs.bib}

\nolinenumbers

\end{document}